\documentstyle[aps,eqsecnum,amsfonts,preprint]{revtex}
\begin{document}
\draft
%
%
\title{\hfill OKHEP-93-11 REVISED\\
Finite-element quantum electrodynamics. II. Lattice
propagators,
current commutators, and axial-vector anomalies}
\author{Dean Miller\thanks{E-mail: dfmiller@phyast.nhn.uoknor.edu},
  Kimball A. Milton\thanks{E-mail:
milton@phyast.nhn.uoknor.edu},
 and Stephan Siegemund-Broka\thanks{E-mail:
sieg@sgs.es.hac.com}}
\address{Department of Physics and Astronomy,
	University of Oklahoma,
	Norman, OK 73019}
\date{\today}
\maketitle
\begin{abstract}
We apply the finite-element lattice equations of motion for quantum
electrodynamics given in the first paper in this series to examine
anomalies in the current operators.  By taking explicit lattice
divergences of the vector and axial-vector currents we compute
the vector and axial-vector anomalies in two and four dimensions.
We examine anomalous commutators of the currents to compute divergent
and finite Schwinger terms.  And, using free lattice propagators,
we compute the vacuum polarization in two dimensions and hence
the anomaly in the Schwinger model.  A discussion of our choice of
gauge-invariant current is provided.
\end{abstract}


\section{Introduction}
\label{sec:intro}

This paper summarizes the status of the finite-element approach to
gauge theories, in which the Heisenberg operator equations of motion
are converted to operator difference equations consistent with
unitarity.
For a review of the entire program, from quantum mechanics to quantum
field theory, see Ref.~\cite{review}.
The gauging of an Abelian theory on a finite-element lattice was
first
carried out in Ref.~\cite{qed}, while the generalization to a
non-Abelian theory was given in Ref.~\cite{nagt}.
The beginnings of constructing a canonical
quantum electrodynamics along these lines were given in
Ref.\cite{feagt1}.
In particular, we showed there that, at least for background fields,
the Dirac equation is {\it unitary}, in the sense that canonical
anticommutation
relations hold at each lattice time.

In this paper, we will first show, in Sec.\ \ref{sec:2},  how to
construct free lattice propagators for electrons and photons.
We remind the reader how interactions are introduced, in Sec.\
\ref{sec:3},
and then, using the gauge-invariant current, compute directly vector
and axial-vector anomalies.  Another anomalous aspect of the
current operators is examined in Sec.\ \ref{sec:4}, where equal-time
current commutators are computed, from which the divergent and finite
Schwinger terms are computed.
In Sec.\ \ref{sec:5} a lattice loop calculation is
described. We compute the vacuum polarization in two
dimensions  using the electron propagator found in
Sec.\ \ref{sec:2}, and obtain results consistent with the known
continuum
Schwinger model.  The continuing direction of our program is
sketched in the concluding Sec.\ \ref{concl}.
In the Appendix we discuss a possible alternative definition  of
the current possible
in Euclidean space, and show why such a current is unacceptable.

\section{Free Dirac Equation.  Lattice Propagators}
\label{sec:2}

We begin by reminding the reader of the form of the free
finite-element
lattice Dirac equation\footnote{Note that we use a different sign of
$\mu$ than in Ref.~\cite{feagt1}, and correspondingly, a different
definition of spinors in (\ref{spinors}) and a change of sign in the
unitary
time evolution operator $U$, (\ref{timeevolution}).}:
\begin{equation}
{i\gamma^0\over h}(\psi_{\overline{\bf m},n+1}-\psi_{\overline{\bf
m},n})
+{i\gamma^j\over\Delta}(\psi_{m_j+1,\overline{\bf
m}_\perp,\overline{n}}
-\psi_{m_j,\overline{\bf
m}_\perp,\overline{n}})-\mu\psi_{\overline{\bf m},
\overline{n}}=0.
\end{equation}
Here $\mu$ is the electron mass, $h$ is the temporal lattice spacing,
$\Delta$ is the spatial lattice spacing, $\bf m$ represents a spatial
lattice
coordinate, $n$ a temporal coordinate,
 and overbars signify forward averaging:
\begin{equation}
x_{\overline m}={1\over2}(x_{m+1}+x_m).
\end{equation}
By transforming this into momentum space we find that the momentum
expansion
of the canonical Dirac field is
\begin{equation}
\psi_{\overline{\bf m},n}=\sum_{s, {\bf
p}}\sqrt{\mu\over\omega}
(b_{{\bf p},s}u_{{\bf p},s}\lambda^{-n}e^{i{\bf p\cdot m}2\pi/ M}
+d^\dagger_{{\bf p},s}v_{{\bf p},s}\lambda^n e^{-i{\bf p\cdot m}2\pi/
M}),
\label{fourier}
\end{equation}
where the spinors are normalized according to
\begin{equation}
\sum_s\tilde u_\pm^{\vphantom{\dagger}} \tilde
u_\pm^\dagger\gamma^0=\mp  {\mu\pm\gamma\cdot\tilde p
\over2\mu}\equiv\pm\Lambda_\pm,
\label{spinors}
\end{equation}
where $u=\tilde u_-$ and $v=\tilde u_+$,
and where
\begin{equation}
\tilde{\bf p}={2{\bf t}\over\Delta},
\quad \omega=\tilde p^0=\sqrt{\tilde {\bf p}^2+\mu^2},\quad
({\bf t_p})_i=\tan p_i\pi/M.
\label{tp}
\end{equation}
In (\ref{fourier}) $\lambda$ is the eigenvalue of the Dirac transfer
matrix
\begin{equation}
\lambda={1+ih\omega/2\over1-ih\omega/2}\equiv
e^{i\Omega(h)h}.
\label{lambda}
\end{equation}
Notice that we may solve (\ref{lambda}) for $\omega$:
\begin{equation}
\omega={2\over h}\tan{h\Omega\over2}.
\label{omega}
\end{equation}
We take $M$, the number of lattice points in a given spatial
direction, to be odd, so that $\psi$ is periodic on the spatial
lattice.  It follows from (\ref{fourier}) that
 the canonical lattice anticommutation relations for the Dirac
fields
\begin{equation}
\{\psi_{\overline{\bf m}^{\vphantom{\prime}},n}^{\vphantom{\dagger}},
\psi_{\overline{\bf m}',n}^\dagger\}
={1\over\Delta^3}\delta_{{\bf m}, {\bf m}'}
\label{anticomm}
\end{equation}
are satisfied if ($L=M\Delta$, the length of the spatial lattice)
\begin{mathletters}
\begin{eqnarray}
\{b^{\vphantom{\dagger}}_{{\bf p},s},
b_{{\bf p}',s'}^\dagger\}&=&{1\over
L^3}\delta_{
\bf p,p'}\delta_{s,s'},\\
\{d^{\vphantom{\dagger}}_{{\bf p},s},
d_{{\bf p}',s'}^\dagger\}&=&{1\over L^3}\delta_{
\bf p,p'}\delta_{s,s'},
\end{eqnarray}
\end{mathletters}
and all other anticommutators of these operators vanish.

At this point we can construct a free lattice Dirac Green's function
according to
\begin{eqnarray}
i G_{{\bf m},n;{\bf m}',n'}&=&\langle
T(\psi^{\vphantom{\dagger}}_{\overline{\bf m},n}
\overline{\psi}_{\overline{\bf m}',n'})\rangle\nonumber\\
&=&\langle\psi^{\vphantom{\dagger}}_{\overline{\bf m},n}
\overline{\psi}_{\overline{\bf m}',n'}\rangle\eta(n-n')-
\langle\overline{\psi}_{\overline{\bf m}',n'}
\psi^{\vphantom{\dagger}}_{\overline{\bf m},n}\rangle\eta(n'-n),
\label{definegreen}
\end{eqnarray}
where the step function is defined by\footnote{Inclusion of the
$x=0$ value of the step function eliminates the following nonlocal
term from (\ref{diracgreen}):
$$\delta_{n,n'}{h\over2 L^3}\sum_{\bf p}e^{i\bf{p\cdot (m-m')}2\pi/M}
(\mu-\bbox{\gamma}\cdot\tilde{\bf p})/\tan h\Omega/2.$$}
\begin{equation}
\eta(x)=\left\{\begin{array}{ll}
1,&x>0,\\
1/2,&x=0,\\
0,&x<0.\end{array}\right.
\end{equation}
Using the fact that $b$ and $d$ are annihilation operators, we easily
see
from the Fourier expansion (\ref{fourier}) that the free Green's
function has
the
following momentum expansion:
\begin{eqnarray}
i G_{{\bf m},n;{\bf m}',n'}={1\over L^3}\sum_{\bf
p}{\mu\over\omega}\big[
&&\Lambda_+\lambda^{n'-n}e^{i{\bf
p\cdot(m-m')}2\pi/M}\eta(n-n')\nonumber\\
\quad\qquad\mbox{}+&&\Lambda_-\lambda^{n-n'}e^{-i{\bf
p\cdot(m-m')}2\pi/M}\eta(n'-n)\big].
\label{defgreen}
\end{eqnarray}
To recast (\ref{defgreen}) in
four-dimensional momentum space, we can use the
identity for integer $p$
\begin{equation}
\oint_\gamma{dz\over2\pi i}{z^{p-1}\over
z-1}=\left\{\begin{array}{ll}
1,&p\ge1,\\
0,&p\le0,\end{array}\right.
\label{identity}
\end{equation}
 where $\gamma$ is a contour encircling
$0$ and $1$
in a positive sense.
It is then a straightforward exercise to show that, apart from a
contact term\footnote{This contact term is
$${h\over4L^3}\delta_{n,n'}\sum_{\bf p}e^{i{\bf p\cdot(m-m')}2\pi/M}
(\mu-\bbox{\gamma}\cdot\tilde{\bf p}),$$
and serves to convert, in the numerator, $\cos^2 h\Omega/2$ to
$\cos^2 h\hat\Omega/2$.},
\begin{eqnarray}
G_{{\bf m},n;{\bf
m}',n'}&=&{h\over4\pi}\int_{-\pi/h}^{\pi/h}
d\hat\Omega e^{-ih\hat\Omega(n-n')}
{1\over L^3}\sum_{\bf p}e^{i{\bf p\cdot(m-m')}2\pi/M}\nonumber\\
&&\quad\times{\gamma^0\sin
h\hat\Omega+(\mu-\bbox{\gamma}\cdot\tilde{\bf
p})
h\cos^2 h\hat\Omega/2\over\cos h(\Omega-i\epsilon)-\cos
h\hat\Omega}.
\label{diracgreen}
\end{eqnarray}
This expression has the expected continuum limit: as $h\to0$ and $n
h\to t$,
(\ref{diracgreen}) becomes
\begin{equation}
{1\over2\pi}\int_{-\infty}^\infty d\hat\Omega
e^{-i\hat\Omega(t-t')}
{1\over L^3}\sum_{\bf p}e^{i{\bf
p\cdot(m-m')}2\pi/M}{\gamma\cdot\tilde p-\mu
\over{\tilde p}^2+\mu^2-i\epsilon},
\end{equation}
where $\tilde p=(\hat\Omega,\tilde{\bf p})$.

 In a similar way we can work out the free lattice Green's function
for the photon.  We first note, from the lattice versions of the
Maxwell equations in the temporal gauge, $A_0=0$, in part,
\begin{equation}
{\bf E}=\dot{\bf A},\quad \dot{\bf E}+\bbox{\nabla}\times{\bf B}=0,
\end{equation}
namely,
\begin{equation}
{\bf E}_{\overline{\bf m},\overline n}
={1\over h}({\bf A}_{\overline{\bf m},n+1}-{\bf A}_{\overline{\bf
m},n}),
\end{equation}
and
\begin{equation}
{1\over h}[(E_i)_{\overline{\bf m},n+1}-{(E_i)}_{\overline{\bf m},n}]
+{1\over\Delta}\epsilon_{ijk}[(B_k)_{m_j+1, \overline{\bf m}_\perp,
\overline{n}}
-(B_k)_{m_j, \overline{\bf m}_\perp,
\overline{n}}]=0,
\end{equation}
that the momentum-space transfer matrix for the photon is
\begin{equation}
T=e^{i\Omega h}
\label{photontrans}
\end{equation}
where
\begin{equation}
\omega={2t\over\Delta}={2\over h}\tan{h\Omega\over2},
\end{equation}
just as in (\ref{omega}).
[Incidentally, notice that this form of the free photon transfer
matrix establishes unitarity of the free lattice Maxwell equations.]

Now we  compute the photon propagator defined by
\begin{eqnarray}
i D^{ij}_{{\bf m},n;{\bf m}',n'}&=&\langle
T(A^i_{\overline{\bf m},n}
A^{j}_{\overline{\bf m}',n'})\rangle\nonumber\\
&=&\langle A^i_{\overline{\bf m},n}
A^j_{\overline{\bf m}',n'}\rangle\eta(n-n')+
\langle A^j_{\overline{\bf m}',n'}
A^i_{\overline{\bf m},n}\rangle\eta(n'-n),
\label{definephoton}
\end{eqnarray}
which, on use of the canonical commutation relations for the photon
creation and annihilation operators, defined through the momentum
expansion
of the photon fields, (2.17) of Ref.~\cite{feagt1},
\begin{equation}
[a_{\bf k}^i, a_{{\bf k}'}^j]=0,\quad[a_{\bf k}^i, a_{{\bf
k}'}^{j\dagger}]=
\delta_{\bf k, k'}f^{ij}({\bf k}),
\end{equation}
where the transverse projection operator is
\begin{equation}
 f^{ij}({\bf k})=\delta_{ij}-{({\bf t_k})_i({\bf t_k})_j
\over({\bf t_k})^2},
\end{equation}
leads to
\begin{eqnarray}
i D^{ij}_{{\bf m},n;{\bf m}',n'}&=&
\eta(n-n'){1\over L^3}\sum_{{\bf k}}{1\over2\omega_k}e^{-i\Omega
h(n-n')}
e^{i{\bf k\cdot(m-m')}2\pi/M}f^{ij}({\bf k})\nonumber\\
&&\quad\mbox{}+\eta(n'-n){1\over L^3}\sum_{{\bf
k}}{1\over2\omega_k}e^{i\Omega h(n-n')}
e^{-i{\bf k\cdot(m-m')}2\pi/M}f^{ij}({\bf k}).
\label{phoprop}
\end{eqnarray}
Use of the identity (\ref{identity}) leads to the four-dimensional
form, apart from a contact term,
\begin{eqnarray}
D^{ij}_{{\bf m},n;{\bf
m}',n'}&=&{h^2\over4\pi}\int_{-\pi/h}^{\pi/h}
d\hat\Omega e^{-ih\hat\Omega(n-n')}
{1\over L^3}\sum_{\bf k}e^{i{\bf k\cdot(m-m')}2\pi/M}f^{ij}({\bf
k})\nonumber\\
&&\quad\times{\cos^2 h\hat\Omega/2\over\cos h(\Omega-i\epsilon)-\cos
h\hat\Omega},
\label{photongreen}
\end{eqnarray}
completely analogous to the electron Green's function
(\ref{diracgreen}).

\section{ Interactions. Axial-vector Anomaly}
\label{sec:3}
Interactions of an electron with a background electromagnetic field
is given in terms of a transfer matrix $T$:
\begin{equation}
\psi_{n+1}=T_n\psi_n,
\end{equation}
which is to be understood as a matrix equation in $\overline {\bf
m}$.
Explicitly, in the gauge $A^0=0$, \cite{feagt1}
\begin{equation}
T=2U^{-1}-1,\quad
U=1+{ih\mu\gamma^0\over2}-{h\over\Delta}\gamma^0\bbox{\gamma
\cdot{\cal D}},
\label{timeevolution}
\end{equation}
where
\begin{equation}
{\cal D}^j_{\bf m,m'}=-(-1)^{m_j+m_j'}[\epsilon_{m_j,m_j'}\cos\hat
\zeta_{m_j,m'_j}-i\sin\hat\zeta_{m_j,m'_j}]\sec\zeta^{(j)}\,
\delta_{\bf m_\perp,m'_\perp}.
\label{eq:d}
\end{equation}
Here
\begin{equation}
\epsilon_{m,m'}=\left\{\begin{array}{ll}1,&m>m',\\
 0,&m=m',\\
-1,&m<m',\end{array}\right.
\end{equation}
and (the following are local and unaveraged in ${\bf m}_\perp$, $n$)
\begin{mathletters}
\begin{equation}
\zeta_{m_j}={e\Delta\over2}A^j_{\overline{m_j-1}},\quad \zeta^{(j)}=
\sum_{m_j=1}^M\zeta_{m_j},
\end{equation}
and
\begin{equation}
\hat\zeta_{m_j,m_j'}=\sum_{m_j''=1}^M{\rm sgn}\,(m_j''-m_j){\rm
sgn}\,(m_j''-m_j')\zeta_{m_j''},
\end{equation}
\end{mathletters}
$\!\!\!\!\!$ with
\begin{equation}
{\rm sgn}\,(m-m')=\epsilon_{m,m'}-\delta_{m,m'}.
\end{equation}
Because ${\cal D}$ is anti-Hermitian, it follows that $T$ is unitary,
that
is,
that $\phi_{{\bf m},n}=\psi_{\overline{\bf m},n}$ is the canonical
field variable satisfying
the canonical anticommutation relations (\ref{anticomm}).

It is instructive, and very simple, to consider the Schwinger model,
that is the case with dimension $d=2$ and mass $\mu=0$.  Because the
light-cone
aligns with the lattice in that case, we set $h=\Delta$.  Then we
see that the transfer matrix for positive or negative chirality,
that is, eigenvalue of $i\gamma_5=\gamma^0\gamma^1$ equal to $\pm1$,
is
\begin{equation}
T_\pm={1\pm{\cal D}\over1\mp{\cal D}}.
\end{equation}
{}From (\ref{eq:d}) we see that the numerator of $T$ is
\begin{equation}
(1\pm{\cal D})_{m,m'}=[\delta_{m,m'}e^{\pm i\zeta}
\mp(-1)^{m+m'}\epsilon_{m,m'}
e^{-i\epsilon_{m,m'}\hat\zeta_{m,m'}}]\sec\zeta,
\end{equation}
while it is a simple calculation to verify that the inverse of this
operator is
\begin{mathletters}
\begin{eqnarray}
(1+{\cal D})_{m,m'}^{-1}&=&
{1\over2}\left(\delta_{m,m'}+\delta_{m,m'- 1}e^{-
2i\zeta_{m'}}\right),\\
(1-{\cal D})_{m,m'}^{-1}&=&
{1\over2}\left(\delta_{m,m'}+\delta_{m,m'+1}e^{2i\zeta_{m}}\right).
\end{eqnarray}
\end{mathletters}
It is therefore immediate to find
\begin{mathletters}
\begin{eqnarray}
(T_+)_{m,m'}&=&\delta_{m,m'+1}e^{2i\zeta_{m}},\\
(T_-)_{m,m'}&=&\delta_{m+1,m'}e^{-2i\zeta_{m'}},
\end{eqnarray}\label{solution}
\end{mathletters}
$\!\!\!\!\!\!$ which
simply says that the $+$ ($-$) chirality fermions move on the
light-cone
to the right (left), acquiring a phase proportional to the vector
potential. Solution (\ref{solution}) directly implies the chiral
anomaly
in the Schwinger model, as shown in \cite{qed}.

To compute the vector and axial-vector anomalies in $(3+1)$
dimensions,
we will find it useful to expand $T$ in powers of
the
temporal lattice spacing $h$ (but not in the spatial lattice spacing
$\Delta$).
This leads to a very simple form for the transfer matrix:
\begin{equation}
T=1+h\left[-i\mu\gamma^0+{2\over\Delta}\gamma^0
\bbox{\gamma\cdot{\cal D}}
\right]+h^2\left[-{\mu^2\over2}+{2\over\Delta^2}{\cal
D}^2+{2i\over\Delta^2}
\bbox{\sigma\cdot({\cal D}\times{\cal
D}})\right]+O(h^3).
\end{equation}

On the lattice the gauge-invariant current is written in terms
of the all-averaged field
$\Psi$,
\begin{equation}
\Psi_{{\bf m}, n}=\psi_{\overline{\bf m},
\overline{n}}={1\over2}(\phi_{{\bf m},n+1}+\phi_{{\bf
m},n}),
\end{equation}
rather than the canonical field $\phi$.
That is, the current is
\begin{equation}
J^\mu_{{\bf m},n}=e\Psi_{{\bf
m},n}^\dagger\gamma^0\gamma^\mu\Psi_{{\bf m},n}^{\vphantom{\dagger}}
={e\over4}[\phi^\dagger_{n}(1+T^\dagger_n)]_{\bf m}
\gamma^0\gamma^\mu[(1+T^{\vphantom{\dagger}}_n)
\phi^{\vphantom{\dagger}}_n]_{\bf m}.
\label{current}
\end{equation}
(For a discussion of why this choice of current is
used, see the Appendix.)
In the absence of interactions it is easy to show that
\begin{equation}
\langle\text{``}\partial_\mu
J^\mu\text{''}\rangle=0,
\end{equation}
where the quotation marks signify a finite-element lattice
divergence.
At the initial time $n=0$ we introduce the momentum expansion
(\ref{fourier}), and compute the expectation value in the
corresponding Fock-space vacuum.
 If we use the expression for
the current
in (\ref{current}) and keep only terms of $O(h)$, we find
\begin{equation}
\langle J^\mu_{\overline{\bf m},1}\rangle=
\langle J^\mu_{\overline{\bf m},0}\rangle-\sum_{\bf m'}{h\over\Delta}
\langle\phi^\dagger_{{\bf m}',0}\gamma^0\bbox{\gamma}\cdot{1\over2}
(\bbox{\cal D}_{(0)}+\bbox{\cal D}_{(1)})^{\vphantom{\dagger}}
_{\bf m',m}\gamma^0\gamma^\mu
\phi^{\vphantom{dagger}}_{{\bf m},0}\rangle+\text{h.c.}
\end{equation}
Thus, to leading order in $h$, the vacuum expectation value of the
divergence of the current is (averaging in ${\bf m}$ is implicit)
\begin{eqnarray}
\langle\hbox{``}\partial_\mu J^\mu\hbox{''}\rangle&=&{\langle
J^0_{\overline{\bf m},1}\rangle
-\langle J^0_{\overline{\bf m},0}\rangle\over h}\nonumber\\
&=&-{2e\over\Delta L^3}\sum_{\bf m'}\sum_{\bf p}{\tilde p_j\over
\omega}
\sin [p_j(m_j'-m_j)2\pi/M]\tilde\alpha^{(j)}_{m_j,m'_j},
\label{vectoranomaly}
\end{eqnarray}
where  $\tilde\alpha=-\text{Im}\, {\cal D}$.
Not only is this not zero, it is not gauge invariant, for under a
gauge transformation,
\begin{equation}
\delta {\cal D}_{m_j,m'_j}^j=ie(\delta\Omega_{m_j}-\delta\Omega_{m'_j})
{\cal D}_{m_j,m'_j}^j.
\end{equation}
This seems to be a conundrum, because the current was explicitly
constructed to be gauge invariant.  The resolution lies in the
recognition that since the canonical field $\phi$ is
covariant under time-independent gauge transformations,
\begin{equation}
\delta\phi_{{\bf m},n}=ie\delta\Omega_{\bf m}\phi_{{\bf m},n},
\end{equation}
 the states created by that operator must transform likewise.  That
is, we require
\begin{equation}
\delta\langle\phi^\dagger_{m_j}\phi^{\vphantom\dagger}_{m_j'}\rangle
=ie(\delta\Omega_{m_j'}-\delta\Omega_{m_j})\langle
\phi^\dagger_{m_j}\phi^{\vphantom\dagger}_{m_j'}\rangle.
\end{equation}
This suggests we supply the following factor in computing the
vacuum expectation values:
\begin{equation}
\langle
\phi^\dagger_{m_j}\phi^{\vphantom\dagger}_{m_j'}\rangle\to
\exp(i\hat\zeta_{m_j,m_j'}\epsilon_{m_j,m_j'})
\langle
\phi^\dagger_{m_j}\phi^{\vphantom\dagger}_{m_j'}\rangle,
\end{equation}
because
\begin{equation}
\delta\hat\zeta_{m_j,m_j'}=e(\delta\Omega_{m_j}-\delta\Omega_{m_j'})
\epsilon_{m_j',m_j}.
\end{equation}
Then, the derivative operator that occurs in (\ref{vectoranomaly})
becomes
\begin{equation}
-\text{Im}\,e^{i\hat\zeta_{m_j,m_j'}\epsilon_{m_j,m_j'}}
{\cal D}^j_{m_j,m_j'}=0,
\end{equation}
since the term with $m_j=m_j'$ does not contribute.

  In the next order in $h$ it appears that a nontrivial contribution
could emerge.
It is the following gauge-invariant structure (here only the the leading
term in $\Delta A$ is written down):
\begin{equation}
\langle\hbox{``}\partial_\mu J^\mu\hbox{''}\rangle\propto
-{2eh\over\Delta^5}
M\sum_i{\zeta^{(i)}}^2.
\end{equation}
This is a nonlocal term without a continuum analogue, because its
gauge invariance depends on the periodic boundary conditions on the
spatial lattice. However, to this order, since $A_{{\bf m},1}=A_{
{\bf m},0}+O(h)$, a simple inspection shows that the coefficient of this
term is zero.

The axial current is defined just as in (\ref{current}) with the
additional
factor  of $i\gamma_5$.
We compute the divergence analogously to (\ref{vectoranomaly}),
except here the first apparently nonzero term is $O(h^2)$:
\begin{eqnarray}
\langle\text{``}\partial_\mu J_5^\mu\text{''}\rangle&=&{\langle
J^0_{5,\overline{\bf m},1}\rangle
-\langle J^0_{5,\overline{\bf m},0}\rangle\over h}\nonumber\\
&\propto& -{2eh\over\Delta^2 L^3}\sum_{\bf m',m''}\sum_{\bf p}{\tilde p_j
\over\omega}\epsilon_{jkl}
\sin [p_j(m_j'-m_j'')2\pi/M]
\tilde\alpha^{(k)}_{m^{\vphantom{\prime}}_k,m'_k}
\tilde\alpha^{(l)}_{m^{\vphantom{\prime\prime}}_l,m''_l}.
\label{axialanomaly}
\end{eqnarray}
Because of the diagonal property of $\cal D$ in the perpendicular
coordinates [see (\ref{eq:d})], (\ref{axialanomaly}) is zero.
So, to $O(h)$ both the vector and axial-vector currents
are nonanomalous.  This may be a bit surprising since when $h=\Delta$
an axial-vector anomaly emerges in $(1+1)$ dimensions.  We are presently
trying to understand
  this state of affairs by carrying out
 an exact calculation with $h=\Delta$
in $(3+1)$ dimensions.

\section{Anomalous Current Commutators}
\label{sec:4}
It is extremely interesting to compute commutators of the current
defined
by (\ref{current}), and compare with the anomalous commutators in the
continuum\cite{Schwinger,BJL}:
\begin{equation}
\langle[J^0(0,{\bf x}),{\bf J}(0)]\rangle=iS\bbox{\nabla}\delta({\bf
x})
+{id}\bbox{\nabla}\nabla^2\delta({\bf x}),
\end{equation}
where $S$ is the quadratically divergent Schwinger term, and
$d=1/12\pi^2$
for the
Bjorken-Johnson-Low regularization.
Straightfoward evaluation of the current-current commutators on the
lattice
requires use of the transfer matrix $T$ for the free Dirac field:
\begin{eqnarray}
T&=&\left({i\gamma^0\over h}+{\bbox{\gamma}
\cdot{\bf t}\over\Delta}-{\mu\over2}\right)^{-1}
\left({i\gamma^0\over h}-{\bbox{\gamma}
\cdot{\bf t}\over\Delta}+{\mu\over2}\right)\nonumber\\
&=&\left(1+{\mu^2h^2\over4}+{h^2\over\Delta^2}t^2\right)^{-1}
\left(1-{\mu^2h^2\over4}
-{h^2\over\Delta^2}t^2
+{2h\over\Delta}i\gamma^0\bbox{\gamma}\cdot{\bf t}
-\mu hi\gamma^0\right),
\label{tmatrix}
\end{eqnarray}
where
${\bf t}={\bf t}_{\bf p}$ is given by (\ref{tp}).
The result is the following expression for a hypercubic lattice,
$h=\Delta$:
\begin{equation}
\langle[J^0_{{\bf m},n},J^j_{{\bf m'},n}]\rangle
=-{4e^2\over (M\Delta)^6}\sum_{\bf k,k'}
{\tan(\pi k'_j/M)\over(1+(\Delta\omega_{\bf
k-k'}/2)^2)(1+(\Delta\omega_{\bf k'}
/2)^2)(\Delta\omega_{\bf k'}/2)}e^{2\pi i{\bf
k\cdot(m-m')}/M}.
\label{comm}
\end{equation}
This matrix (in $\bf m$ and $\bf m'$) is shown for lattice size $M=9$
and mass $\mu=0$ in Fig.\ \ref{fig1}.
The abscissas of the plot correspond to one-dimensional
representations of the
three components of $\bf m$ and $\bf m'$, respectively.  (That is,
the base-$M$ number $(m_1,m_2,m_3)$ is converted to a base-10 value
of the
abscissa.)  It is apparent that the leading-order behavior of this
commutator is a first derivative of the delta function, as expected.

To make this comparison quantitative, we fit this result to the
functional
form expected in the continuum.  The lattice analog of the Dirac
delta function is
\begin{equation}
\delta({\bf x})\approx{1\over\Delta^3}\delta_{\bf m,0}={1\over
(M\Delta)^3}
\sum_{\bf k}e^{2\pi i{\bf k\cdot m}/M},
\end{equation}
so we take as a trial function
\begin{equation}
-e^2{1\over(M\Delta)^3}{2\pi i\over M\Delta}\sum_{r=0}^R
a_{2r+1}P_r(\Delta)(-1)^r\left({2\pi\over M\Delta}\right)^{2r}
\sum_{\bf k} k_jk^{2r}
e^{2\pi i{\bf k\cdot (m-m')}/M}.
\label{fit}
\end{equation}
When $\mu=0$,
the coefficients $a$ do not depend on the lattice spacing $\Delta$;
they
will presumably remain finite in the continuum limit, $\Delta\to0$.
The functions $P_r(\Delta)$ are inserted to force each term in the
series
to be of the same order in $\Delta$ as the commutator (\ref{comm}),
namely
$1/\Delta^6$:
\begin{equation}
P_r(\Delta)=\Delta^{2r-2}.
\end{equation}
 Therefore, in the continuum
limit, the first term in the series (\ref{fit}) will be quadratically
divergent,
the second will be finite, and the rest will vanish, as expected.

To perform the fit, the problem is first converted from a
three-dimensional
 to a two-dimensional
representation.  (This is necessary because the generation of
three-dimensional plots
is impractical beyond $M=9$ and three-dimensional
 fits are prone to difficulties.)
Specifically, a plot such as Fig.\ \ref{fig1} is projected
onto a plane orthogonal to the $m$, $m'$ plane and to the diagonal of
the
matrix.    The resulting
curve
is then fit to a similar projection of (\ref{fit}).
  In Fig.\ \ref{fig4} we show the results for the
coefficients
of a  fit including only the first three odd derivatives
of the lattice delta function
(that is, for $R=2$ in (\ref{fit}))
 for various lattice sizes.
The coefficient of the first derivative is roughly consistent with
the
Schwinger
result\cite{Schwinger}. Similarly, the result for the
coefficient of the third derivative term in the commutator is in
approximate agreement with the
agreement with the BJL  result \cite{BJL}; but seems to be closer
to that of Chanowitz \cite{chanowitz}.
It is not surprising that the results are regularization dependent,
given the divergent nature of the commutator.
A more detailed discussion of this calculation will appear elsewhere
\cite{miller}.

\section{Loop Calculation. Schwinger Model}
\label{sec:5}

It is illuminating to calculate physical quantities such as the
magnetic
moment or the axial-vector anomaly by doing perturbation theory on
the lattice.
That is, we calculate quantities such as vacuum polarization by using
weak-coupling perturbation theory together with the
lattice propagators
 given in (\ref{diracgreen}) and (\ref{photongreen}).
We illustrate this idea in the simple context
of the
Schwinger model, two-dimensional QED.  (An analogous calculation in
four
dimensions of both the axial-vector anomaly and the anomalous
magnetic
moment is in progress.)

We begin by discussing the Schwinger model  ($d=2$ electrodynamics)
in the continuum.
To calculate the anomaly it is essential to work with massive QED$_2$
and then take the massless limit to avoid infrared singularities.
The vacuum polarization is then
\begin{equation}
\Pi^{\mu\nu} = ie^2 \int {d^2p\over (2\pi)^2}
{{\rm tr} \left(\gamma^\mu ({p \kern -.50em /}-{k \kern -.50em
/}-\mu) \gamma^\nu
({p \kern -.50em /}-\mu)\right)
\over ((p-k)^2+\mu^2)(p^2+\mu^2)}.
\end{equation}
We require that $\Pi^{\mu\nu}$  be transverse:
\begin{equation}
\Pi^{\mu\nu} = \Pi(k^2)(g_{\mu\nu}-{k_\mu k_\nu \over k^2}).
\end{equation}
Contracting on the indices gives
\begin{equation}
\Pi(k^2) = -ie^2 \int {d^2p\over (2\pi)^2} {4\mu^2 \over
 ((p-k)^2+\mu^2)(p^2+\mu^2)},
\label{vp}
\end{equation}
which is easily expressed using Feynman parameters as
\begin{equation}
\Pi(k^2)=-{e^2\over \pi}\int_0^1 dx{\mu^2\over
x(x-1)k^2+\mu^2}.
\end{equation}
  For the $\mu\ne0$ theory, we require that $\Pi(0)=0$.  This implies
the presence of a subtraction constant $e^2/\pi$, which for the
$\mu=0$
theory is the square of the boson mass, or the anomaly.
Rather than introduce a Feynman parameter we could obtain this result
by setting $k=0$ in (\ref{vp}) and performing the
momentum integration explicitly.  We have, then, for the unsubtracted
vacuum polarization
\begin{equation}
\Pi(k=0)= -i{e^2\over (2\pi)^2}\int dp_0 \,dp_1
{4\mu^2\over (p_0^2-p_1^2 -\mu^2 +i\epsilon)^2}
= -{e^2 \mu^2\over \pi}\int^\infty_0 dp_1 {1\over
(p_1^2+\mu^2)^{3\over 2}},
\end{equation}
which again gives the usual result, $\Pi(0)=-e^2/\pi$.

We now describe the analogous calculation in the finite-element
lattice
framework.  The lowest-order vacuum polarization is given by
 the time-ordered products
of currents,
\begin{equation}
\Pi^{\mu\nu}_{m,n;m',n'} = i\langle J^\mu_{m,n} J^\nu_{m',n'} \rangle
             = ie^2\langle {\overline\Psi}_{m,n}\gamma^\mu \Psi_{m,n}
                       {\overline\Psi}_{m',n'}
\gamma^\nu \Psi_{m',n'} \rangle,
\end{equation}
where the gauge invariant electromagnetic current is given by
 (\ref{current}).
Using the lattice Green's function (\ref{diracgreen})
and the transfer matrix (\ref{tmatrix}),
we find
\begin{eqnarray}
&&\qquad\Pi^{\mu\nu}_{m,n;m',n'}=
ie^2\left({ h\over L 4\pi}\right)^2 \sum_{{ p}_1}\sum_{ p_2}
\int_0^{2\pi/h}\int_0^{2\pi/h}d{\hat\Omega}_1d{\hat\Omega}_2
e^{ih{\hat\Omega}_1(n-n')}e^{-ih{\hat\Omega}_2(n-n')}\nonumber\\
&&\quad\times
e^{-i{ p}_1({ m}- { m'})2\pi/M}
e^{i{ p}_2({ m}- { m'})2\pi/M}
 {\rm tr}\big[ \gamma^{\mu}{{\cal P}}_1
\left(\gamma^0 \sin(h{\hat\Omega}_1) +(\mu-{\gamma}
\tilde{ p}_1)h\cos^2(h\hat\Omega_1/2)\right)\nonumber\\
&&\quad\times
\overline{\cal P}_1 \gamma^{\nu} {{\cal P}}_2
\left(\gamma^0 \sin(h{\hat\Omega}_2) +(\mu-{\gamma}
\tilde{ p}_2)h\cos^2(h\hat\Omega_2/2)\right)
\overline{\cal P}_2  \big]\nonumber\\
&&\quad\times{1\over(\cos{h(\Omega_1-i\epsilon)}-\cos{h\hat\Omega_1)}
(\cos{h(\Omega_2-i\epsilon)}-\cos{h\hat\Omega_2)}},
\end{eqnarray}
 where ${\cal P}=(1+T)/2$ and $\overline{\cal P}=\gamma^0{\cal
P}^\dagger
\gamma^0$.
We now perform a lattice Fourier transform:
\begin{equation}
\Pi^{\mu\nu}(q,\omega)
=
\Delta\sum_{m=1}^M\sum_{n=-\infty}^\infty{2\over{\omega}}\sin{\omega
h\over2}
e^{-ih\omega (n-n')}
e^{i{ q}{(m-m')}2\pi/M}
\Pi^{\mu\nu}_{ m,n;m',n'}.
\end{equation}
Note that
\begin{equation}
\sum_{m=1}^M e^{i{ q}{ m}2\pi/M}= M\delta_{ q,0}
\quad{\rm and}\quad
\sum_{n=-\infty}^\infty e^{-ih\omega n} = 2\pi \delta(h\omega),\quad
-\pi<h\omega\le\pi.
\end{equation}
Upon contracting with $g_{\mu\nu}$ we find for the vacuum
polarization
\begin{eqnarray}
&&\qquad\Pi(q,\omega)=-ie^2\left({ h\over L 4\pi}\right)^2 2\pi
\Delta M
\left({2\over h{\omega}}\sin{\omega h\over2}\right)
\nonumber \\
&&\times\sum_{{ p}_2}\int_0^{2\pi/h}{d{\hat\Omega}_2\,
16\nu^2 r^2 \cos^2(h\hat\Omega_1/2) \cos^2(h\hat\Omega_2/2) \over
(\cos{h(\Omega_1-i\epsilon)}-\cos{h\hat\Omega_1)}
(\cos{h(\Omega_2-i\epsilon)}-\cos{h\hat\Omega_2)}}\nonumber\\
&&\qquad\times{1\over(1+r^2\nu^2+r^2t_1^2) (1+r^2\nu^2+r^2t_2^2)},
\end{eqnarray}
where $\nu = {1\over2}\mu\Delta$ and $r=h/\Delta$ and where, in the
trace,
we recall that $\gamma^\lambda\gamma^\alpha\gamma_\lambda=0$ in two
dimensions.
Here we understand that $\hat\Omega_1 = \omega+\hat\Omega_2$
and ${t}_1 = {t}_{p_1}={t}_{q+p_2}$.
The integral over $\hat\Omega_2$ can now be done to yield, when
${t}_q=\omega=0$,
\begin{equation}
\Pi(0,0)=
-{e^2\over \pi}\sum_{{ p}_2}4\pi{r^3\over M}
{\nu^2 \cos^4{h\Omega_2/2} \over (1+r^2\nu^2+r^2t_2^2)^2}
{\cos{h\Omega_2}\over\sin^3{h\Omega_2}}
\left(1-{4\sin^2h\Omega_2/2\over
\cos h\Omega_2}\right).
\label{sum}
\end{equation}
For zero external momentum, $\Omega_1=\Omega_2$,
where
\begin{equation}
h\Omega_2 = 2\tan^{-1}(r(t^2_2+\nu^2)^{1\over2}).
\end{equation}

We perform the momentum sum in (\ref{sum})
 for various values of the lattice size $M$,
mass parameter $\nu$, and ratio of temporal and spatial  lattice
spacing $r$.
The parameter $\nu$ is constrained by $M^{-1}\ll\nu\ll1$.
Figure \ref{fig5} is a graph of the sum for $M=2533$ versus $\nu$.
Plotted
are
curves with $r= 0.1$, $0.75$, $0.99$.  The first two cases stay close
to $1$ for a significant range of $\nu$, and
only for $r = 0.99$ does the value drop significantly below $1$.
Note
that
for very small $\nu$ the sum diverges, as one expects for
$\nu\sim M^{-1}$.  This feature is shown more explicitly in Fig.\
\ref{fig6},
that is a plot for $r = 0.1$ for lattice sizes $1105\le M\le3145$,
with
increasing endpoint for decreasing lattice size.  For smaller
lattices,
the curve turns up for larger $\nu$, as the limit of the domain of
validity of the lattice calculation is reached.

More detailed analysis shows that, as a function of $r$,
$-\Pi(0,0)/(e^2/\pi)$
is on the average unity, but is very noisy, with wild fluctuations
around 1.  For small $r$,
fluctuations are greatly diminished, but for $r\sim1$
and for various
values of the input parameters, $h\Omega_2$ can get very close to
$\pi$ and make a few terms in the sum very large. (One might think
that  the desired value of $r$ is $r\sim1$, the case of a
hypercubic lattice; as noted in Sec. \ref{sec:3}
the anomaly in the Schwinger model has previously been calculated
with the finite element lattice technique with $r=1$.)
  This question of stability is also under intense scrutiny.

\section{Conclusions}
\label{concl}
This paper is a report of recent progress in our finite-element
program as applied to electrodynamics.  More complete treatments
of the analyses of the last two sections will appear elsewhere
\cite{miller,sb}.  Our immediate goal in QED remains the extraction
of a nonperturbative value for the anomalous moment of the electron.

Of course, our real interest lies in non-Abelian theories.  Our
first task there is to complete the formulation in four dimensions
\cite{nagt,singapore}.  This should be accomplished in the next few
months.  Then we can apply this approach to theories such as QCD.

\acknowledgments
We thank the U.S. Department of Energy for financial support.
DM thanks the U.S. Department of Education for support.  KAM thanks
Fermi National Accelerator Laboratory for its hospitality while
 some of the work reported
here was done.  He also thanks Tai Wu for useful conversations.

\appendix
\section*{Current constructed from Euclidean Lagrangian}
In the text we simply assumed a form of the current (\ref{current})
which was manifestly gauge invariant.  We are at liberty to do
so, because the Minkowski finite-element equations of motion are
not derivable from a Lagrangian.  The current cannot be derived
from the Dirac equation, but is an independent source for the Maxwell
equations, see Ref.~\cite{nagt}.  However, if one were to work in
Euclidean space-time (periodic in all four directions),
it is possible to construct an action
from which the equations of motion are derivable, and which therefore
supplies a lattice current.  The fermion part of that action is
(a factor of $i$ is absorbed in going to Euclidean space)
\begin{equation}
W_f=\Delta^4\sum_{\bf m, m'}\psi^\dagger_{\overline{\bf m}}
\left({2\over \Delta}\bbox{\Gamma\cdot{\cal D}}
+\mu\right)_{\bf m,m'}\psi^{\vphantom\dagger}
_{\overline{\bf m}'},\quad
\bbox{\Gamma}=(\gamma^0\gamma_k,\gamma^0).
\label{action}
\end{equation}
If we vary (\ref{action}) with respect to $\psi^\dagger$ we
obtain the Euclidean Dirac equation
\begin{equation}
\left({2\over \Delta}\bbox{\Gamma\cdot{\cal D}}
+\mu\right)\psi=0.
\label{euclideq}
\end{equation}
where, as in (\ref{action}), a four-dimensional scalar product is
implied.

Given an action, we can construct a conserved vector current
by making a local gauge transformation
\begin{equation}
\delta\psi_{\overline{\bf m}}=
ie\delta\Omega_{\bf m}\psi_{\overline{\bf m}}.
\label{gt}
\end{equation}
Because the Dirac equation, and hence the action, is invariant
under the global version of (\ref{gt}), $\delta\Omega_{\bf m}
=\delta\Omega= \text{constant}$,  we must have by the action
principle
\begin{equation}
\delta W_f=0=-\sum_{\bf m}J^i_{\bf m}
{1\over\Delta}(\delta\Omega_{m_i,{\bf m}_\perp}
-\delta\Omega_{m_i-1,{\bf m}_\perp}),
\end{equation}
from which we read off the  conserved current
\begin{eqnarray}
J^i_{\bf m}=-e&&\sum_{m_i',m_i''}
\psi_{\overline{m}_i',
\overline{\bf m}_\perp}^\dagger \Gamma^i
\psi_{\overline{m}''_i,\overline{\bf m}_\perp}^{\vphantom\dagger}
\,\text{sgn}(m_i-m_i')\,\text{sgn}(m_i-m_i'')\nonumber\\
&&\times(-1)^{m'_i+m_i''}\sec\zeta^{(i)}\exp(-i\epsilon_{m_i',m_i''}
\hat\zeta_{m_i',m_i''}).
\label{eculidcur}
\end{eqnarray}
(The same result, of course, can be obtained by  varying
${\cal D}$ with respect to $A^i_{\overline{m_i-1},{\bf m}_\perp}$.)
The expression for this Euclidean current has been simplified
by deleting constant terms.  It is easy to verify explicitly
that this current is both conserved and gauge invariant.
Similarly, by making a chiral transformation,
\begin{equation}
\delta\psi_{\overline{\bf m}}=\gamma^0\gamma_5\delta\Omega_{\bf m}
\psi_{\overline{\bf m}},
\end{equation}
we can construct the axial-vector current $J^i_{5{\bf m}}$, which
has the form of (\ref{eculidcur}) with the replacement
\begin{equation}
e\Gamma^i\to \gamma^0i\gamma_5\Gamma^i\equiv \Gamma_5^i,\quad
\Gamma_5^i=(-i\gamma_5\gamma^i, -i\gamma_5).
\end{equation}
By construction, these currents possess no anomalies.  However,
they appear to be completely unacceptable, because they are
horribly nonlocal.  In particular, they possess no Minkowski analogues,
in the sense that it is not possible to analytically continue
back to real unbounded times.
Crucial to our formulation is the propagation of the operators
from past times to the present time, so that we can solve for the
field operators by time-stepping through the lattice.
The Euclidean current
(\ref{eculidcur}) involves fermion field operators at all
Euclidean times, which  would make it impossible to solve
for the operators at time $n$ in terms of operators at earlier
times.  Therefore, for the considerations of the
text we use the gauge-invariant current (\ref{current}) and its
axial analogue, currents which do possess anomalies.

\begin{figure}
\caption{Three-dimensional plot of (\protect\ref{comm}) for $M=9$ and
$\mu=0$.}
\label{fig1}
\end{figure}

\begin{figure}
\caption{Coefficients of three spectral fit for different
lattice sizes
($\mu=0$). For the first derivative term, the coefficient shown
is $S\Delta^2$.}
\label{fig4}
\end{figure}

\begin{figure}
\caption{Plot of the sum (\protect\ref{sum}) for $M=2533$ as a
function of
$\nu
=\mu\Delta/2$.  Shown are curves with $r=h/\Delta=0.1$, 0.75, 0.99.}
\label{fig5}
\end{figure}

\begin{figure}
\caption{Plot of (\protect\ref{sum}) for $1105\le M\le3145$ and
$r=0.1$.}
\label{fig6}
\end{figure}

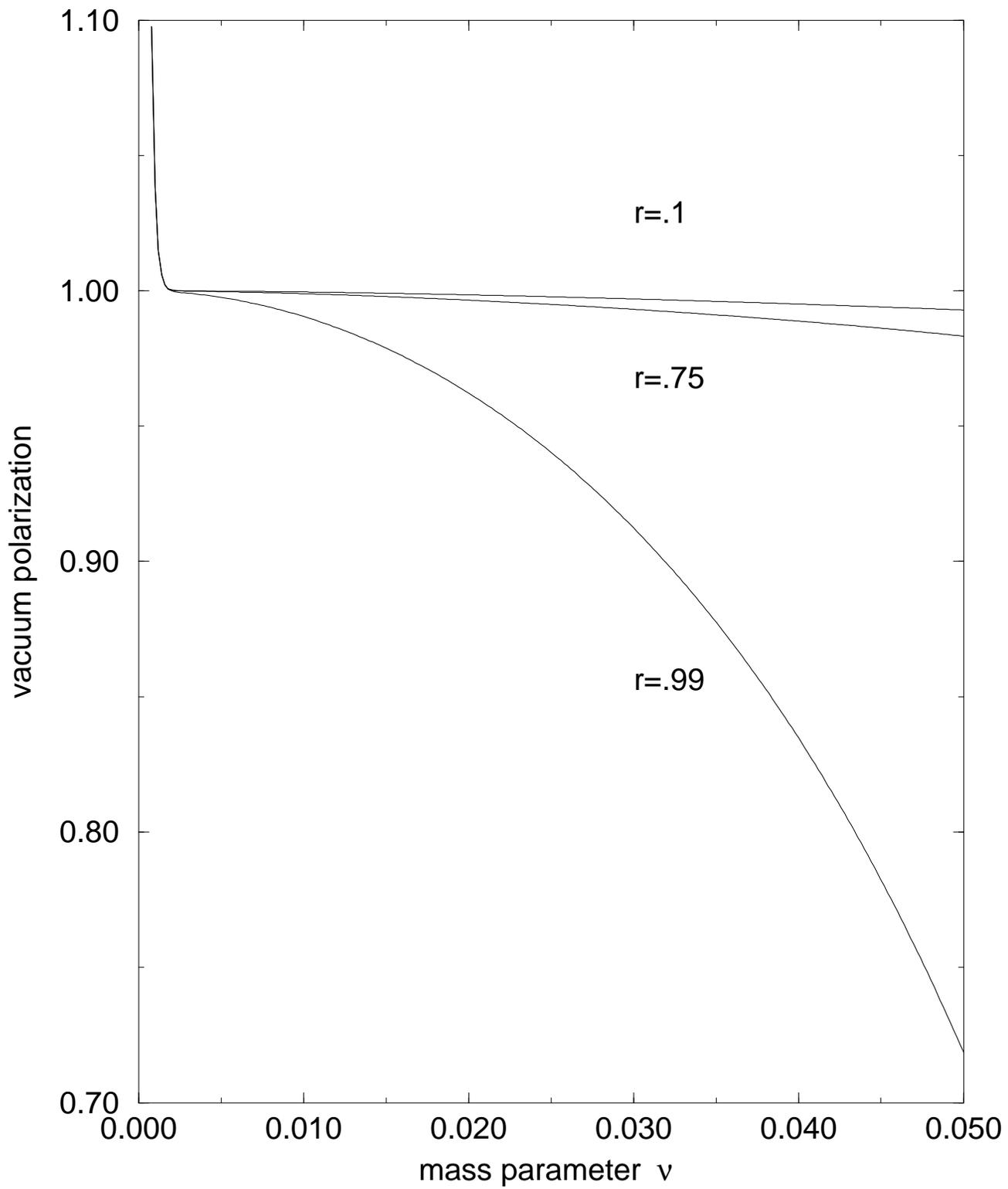
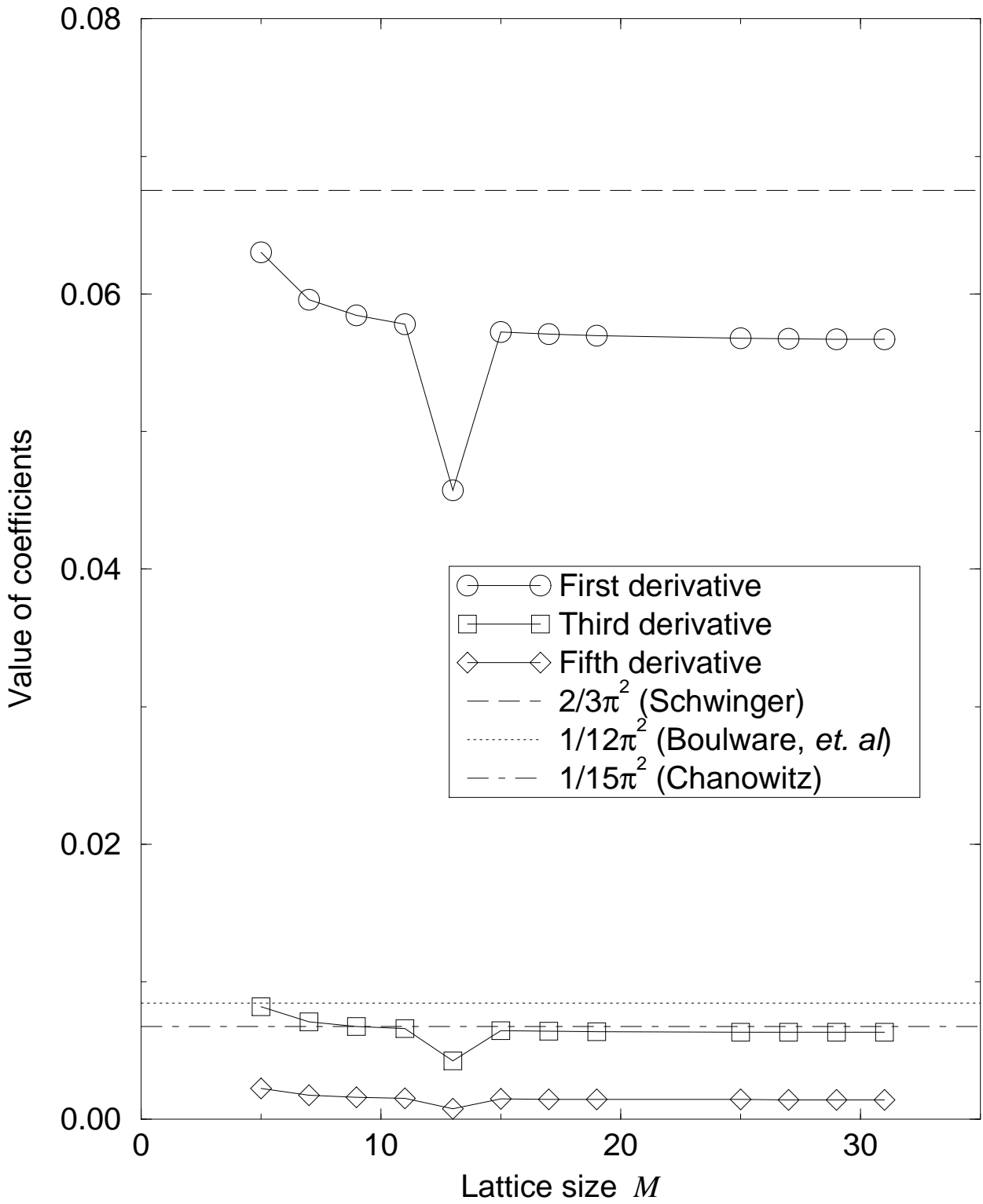
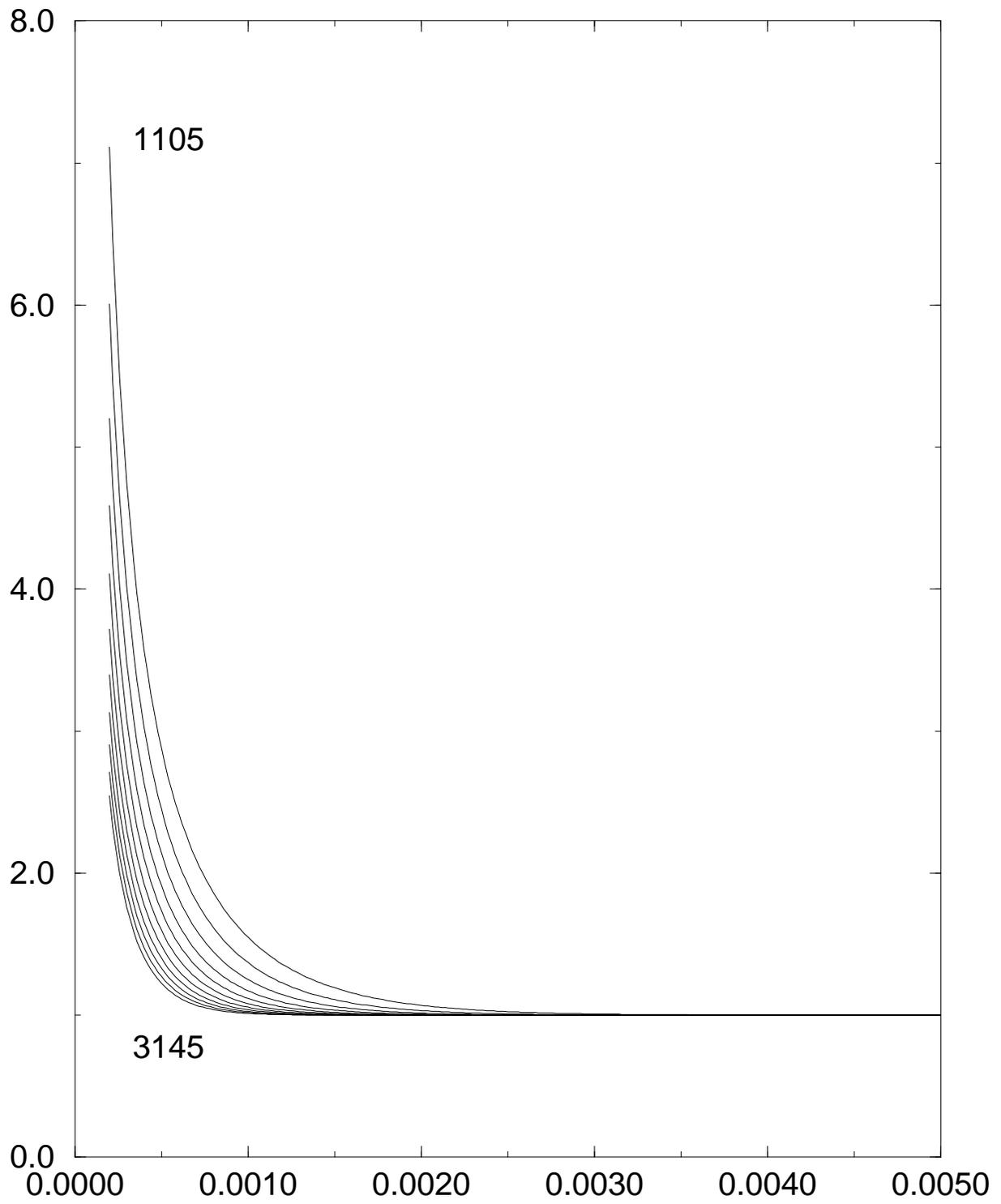
\end{document}